# Scholarly Production and Public Health Determinants in Context of Funding: The Case of IoMT Research:

Peter Kokol, University of Maribor, Faculty of Electrical Engineering and Computer Science, Koroška ulica 46, 2000 Maribor, Slovenia

**Abstract:**

The Internet of Medical Things (IoMT) represents a transformative technology that connects medical devices, sensors, and healthcare systems to enable real-time monitoring, data sharing, and advanced decision-making in healthcare. While the technical and clinical potential of IoMT has been researched extensively, the scale and scope of research funding and their influence on research literature production patterns and country health determinants remain unknown. The study presented in this paper covers this gap by employing triangulation of quantitative and qualitative approaches. The results reveal a positive trend IoMT in research literature production. Thematic analysis shows that both funded and non-funded are associated with similar themes; however, founded research is more focused on recent research trends like artificial intelligence applications in healthcare. Finally, our study revealed the positive association between the number of funded papers and health determinants, suggesting that IoMT research funding might contribute to improved healthcare delivery.

**Keywords:** Internet of Medical Things, Research funding, Synthetic Knowledge Synthesis

## 1. Introduction

New health paradigms advocate the idea that human health is a public good and that the health system's primary goal is to effectively and efficiently manage this good [1, 2]. To achieve this goal, modern health professionals should benefit from significant advancements in medical technology and treatment methods, including innovations in diagnostics, medical imaging, and preventive health care. Remote health status monitoring and therapeutic techniques enabled by the digital and agile health revolution [3–8]. Researchers and healthcare innovators hope that emerging technologies will significantly enhance the efficiency and capacity of health systems, helping them to better meet the healthcare needs of populations worldwide [9]. One of the promising technologies that can be widely used in healthcare for all of the above purposes is the Internet of Things (IoT) [10]. While the Internet mainly connects people, IoT connects not only people but different entities in the physical world, including people and objects. The IoT has made remarkable progress across many fields, including medicine, where IoT is called the Internet of Medical Things (IoMT) [11–14]. The IoMT refers to the interconnected network of medical devices, sensors, software applications, and healthcare systems that communicate through the internet, enabling real-time data sharing, remote monitoring, and enhanced decision-making [15], supporting the patient-centric approach [16].

Research funding is believed to be a catalyst for the science and technology progress. As global increase in research expenditure has been noted in last years, analysing the effectiveness or return of research funding has become an important topic of interest [17]. In addition, to being an indicator of a research area importance, impact, maturity and visibility, funding information might be used to identify funding possibilities, research patterns, and most probable topics and themes to be funded [18]. While IoMT emergence has been researched from various aspects and viewpoints, the scope and extent of research funding have not yet been analysed. The aim of this paper is to fill this gap and answer the following research questions:

- What are the prevailing research literature production and bibliometric patterns in funded and non-funded publications concerning funded entities, rends, most prolific source titles and most prolific funded research themes?
- Which are the most productive funding agencies and countries associated with IoMT research?
- What is the possible impact of funded research on the country's health determinants?

## 2. Materials and Methods

Scopus (Elsevier, The Netherlands) was used as the bibliographic database due to the fact that Scopus is regarded as the largest scientific bibliographic database of the reviewed research literature. In addition to providing powerful analytics services, Scopus also enables 20,000 records to be exported simultaneously, enabling more effective and efficient bibliometric analyses. Another strength of Scopus compared to other bibliographic databases is its completeness in covering funding information [18, 19]. The search query was formed by analysis and synthesis search strategies used in published IoT/IoMT review papers. The search was performed on August 2$^{nd}$, 2025. Zipf's law was used to calculate the number of relevant keywords to be used in Synthetic Knowledge Synthesis (SKS) [20]. The following search string was used for funded publications (FPs):

TITLE-ABS-KEY("Internet of Medical Things") and FUND-ALL(a* or b* or c* or d* or e* or f* or g* or h* or i* or j* or k* or l* or m* or n* or o* or p* or q* or r* or s* or t* or u* or v* or z* or x* or y* or w* or 1* or 2* or 3* or 4* or 5* or 6* or 7* or 8* or 9* or 0*)

TITLE-ABS-KEY("Internet of Medical Things") and NOT FUND-ALL(a* or b* or c* or d* or e* or f* or g* or h* or i* or j* or k* or l* or m* or n* or o* or p* or q* or r* or s* or t* or u* or v* or z* or x* or y* or w* or 1* or 2* or 3* or 4* or 5* or 6* or 7* or 8* or 9* or 0*) for non-funded publications (NFPs). We excluded the year 2025, due the fact that country determinants data for 2025 are not yet available.

First two research questions were answered using Scopus's built-in functionality. The country ranks in Medicine, and Computer networks and communications research were obtained from Scimago [21]. Country determinants used in our study are Health systems ranking [22], Health Expenditure as % of GDP, Current R&D Expenditure as % of GDP [23] CR&D Expenditure as % of GDP [23] and Bloomberg Global Health Index (BGHI) [24]. BGHI accounts for a variety of factors that contribute to the populational health of countries. It is based on the premise that the residents of developed countries tend to be healthier due to a higher quality of life, lower pollution rates, better infrastructure, access to quality healthcare, education, better nutrition, and good-paying jobs. On the contrary, residents of other developing countries frequently lack adequate access to these same benefits. The associations between bibliometric dimensions and country determinants were analysed by a framework developed by Kokol et al [25–27].

Thematic analysis was performed by Synthetic Knowledge Synthesis (SKS) [28].

## 3. Results

The search resulted in 3943 publications, 3481 NFPs and 453 FPs. The percentage of FPs was 11.5%, which is lower than in related research fields Internet of Things (IoT) (15.8%) and lover then in IoT use in preventive health (37.0%) [10].

### *3.1.Bibliometric patterns of IoMT research and its influence on country determinants*

Figure 1 shows a growth in the number of papers in the domain of IoMT, for both NFPs and FPs, from 2016 to 2024. However, while the growth in NFPs is resembling the exponential curve, the growth in FPs is approximately linear. Initially the ratio of funding papers increased from 12% to 27% in 2019, then the ratio steadily decreased to less than 11%. In 2024.

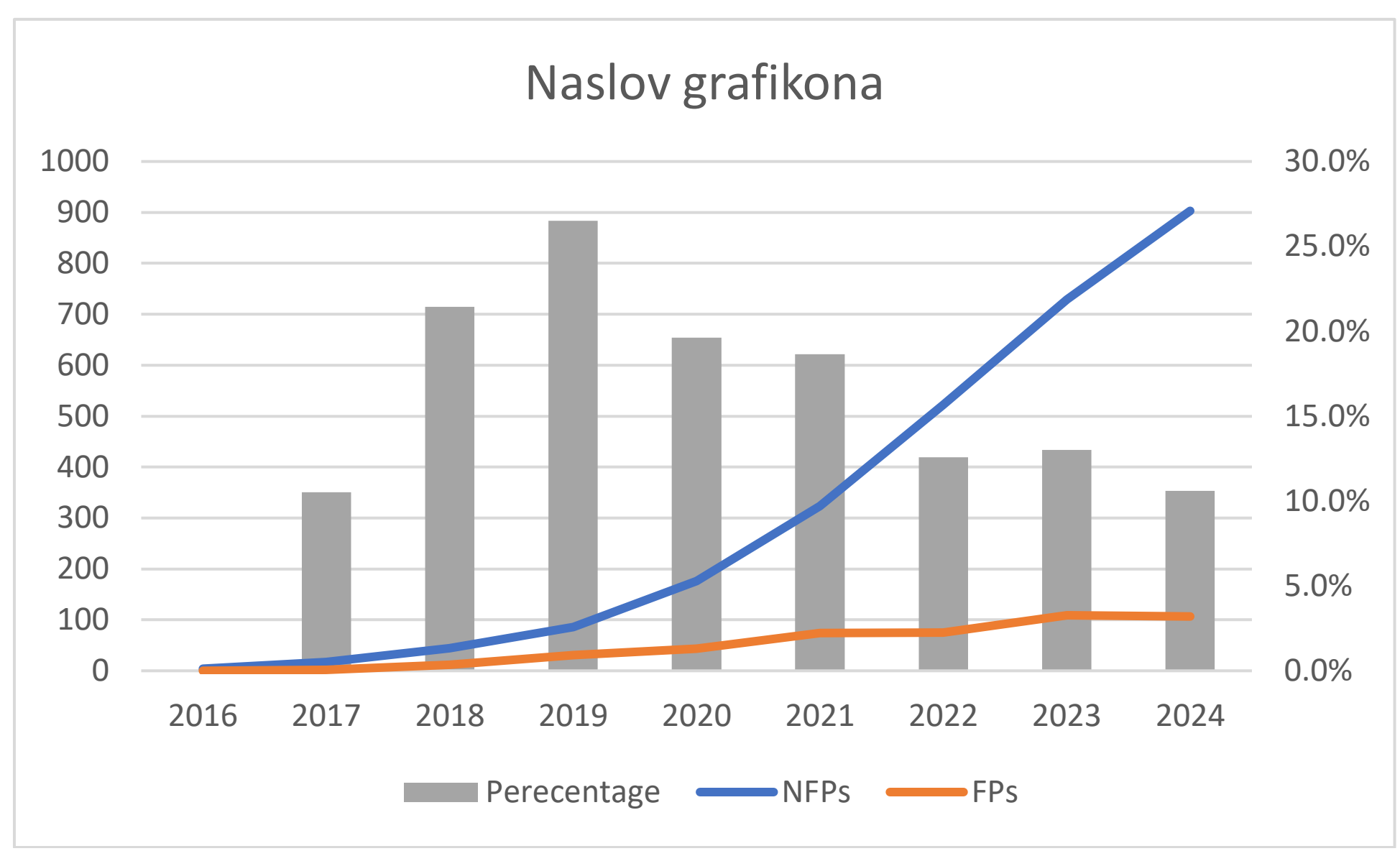

**Figure 1.** The literature production dynamics of FPs and NFPs in IoMT research.

All five most productive source titles publishing funded publications on IoMT are ranked in the first quartile according to Scpus CiteScore, more precisely their average ranks percentil is almost 90% (Table 1).

**Table 1.** Most prolific source titles publishing FPs

| SOURCE TITLE | Number of FPs | CiteScore | Percentil |
|---|---|---|---|
| IEEE Access | 29 | 9.0 | 90 |
| Sensors | 23 | 8.2 | 88 |
| IEEE Internet Of Things Journal | 22 | 16.3 | 97 |
| Future Generation Computer Systems | 13 | 17.1 | 97 |
| Flectronics Swit-zerland | 12 | 6.1 | 75 |

| Average | | 11.34 | 89.4 |
|---|---|---|---|

Four of the five most productive source titles publishing non-funded publications on IoMT are ranked in the first quartile according to Scpus CiteScore (Table 2). Three of the top five journals published both FPs and NFPs. The ranking and CitesCores of both top group journals are compaurable, however NFPs are also published in Lecture notes which is not a regular journal but a proceeding series, with lower CiteScore.

**Table 2.** Most prolific source titles publishing NFPs

| SOURCE TITLE | Number of FPs | CiteScore | Percentil |
|---|---|---|---|
| IEEE Internet Of Things Journal | 142 | 16.3 | 97 |
| IEEE Access | 133 | 9.0 | 90 |
| Lecture Notes In Networks and Systems | 103 | 1.0 | 21 |
| Sensors | 63 | 8.2 | 88 |
| IEEE Journal of Biomedical and Health Informatics | 62 | 13.5 | 95 |
| Average | | 9.6 | 78.2 |

**Table 3** reveals that the country productivity ranks of both NFPs, and FPs differs from the country productivity ranks in subjects Medicine and Computer networks and communications. Three most productive countries in above two subjects are still among 10 top productive countries, but top productive countries like Germany, Japan, France and Canada are missing. On the other hand, countries which don't belong to top 30 most productive countries like Saudi Arabia, Pakistan and Malaysia are among top 10 productive countries in IoMT research. Eight countries are in both NFP and FP list, however the rankings differ slightly, India and United States being the most productive regarding NFPs and China and Saudi Arabia regarding FPs. Australia and Iraq are only ranked among top 10 NFPs countries and South Korea and Egypt only among FPs countries. The average percentage of funded papers in top 10 countries is 44.2% with South Korea having the largest percentage of funded papers (69.3%) and India the smallest percentage (20.0%).
According to Organisation for Economic Co-operation and Development (OECD) [29] the average health spending related to GPD in 2022 in OECD countries is about 9.2%.Thereafter it is surprising to observe that only three countries in the top 10 NFPs or FPs countries are above that limit. Similarly, the most of top ranked countries in Health and Health systems ranking like Singapore, Japan, Taiwan, Scandinavian Countries, etc. are also not in the below lists. It seems that countries with low health expenditures and not so well ranked health systems are investing in the research in IoMT as a possible technology to improve health services. Similar observation can be made regarding the R&D expenditures. According to OECD [23] an appropriate spending in percentages of GPD is above 2% and only four of top 10 countries reached that limit. Data reveals that countries highly above or highly below 2% expenditure in R&D GPD are the most productive regarding both NFPs and FPs. A comparison of country determinants between the most productive countries in the NFP and FP lists reveals that the countries on the FP list rank higher on average in all three rankings, have a higher percentage of investments in healthcare and research and higher Global Health Index.

**Table 3**. Most productive countries and their country determinants

| Funded/Non funded publication | COUNTRY/TERRITORY | Number of NFPs | Scimago rank in subject Medicine | Scimago rank in sub-subject Computer networks and communications | Health systems ranking 2023 [22] | Current Health Expenditure as % of GDP - 2021/22 | Current R&D Expenditure as % of GDP - 2021/22 [23] | BGHI |
|---|---|---|---|---|---|---|---|---|
| NFP | India | 849 | 11 | 3 | 112 | 3.28 | 0.65 | 61.3 |
| NFP | United States | 213 | 1 | 2 | 69 | 16.57 | 3.45 | 79.5 |
| NFP | China | 192 | 2 | 1 | 5 | 5.38 | 2.43 | 46.3 |
| NFP | Saudi Arabia | 145 | 35 | 27 | 56 | 5.97 | 0.46 | 77.2 |
| NFP | Pakistan | 112 | 40 | 30 | 124 | 2.91 | 0.16 | 61.5 |
| NFP | United Kingdom | 93 | 3 | 6 | 34 | 11.34 | 2.91 | 88.8 |
| NFP | Australia | 73 | 9 | 13 | 21 | 10.54 | 3.25 | 90.9 |
| NFP | Iraq | 69 | 65 | 50 | 115 | 5.25 | 0.04 | 62.8 |
| NFP | Italy | 63 | 6 | 8 | 17 | 9.00 | 1.45 | 91.5 |
| NFP | Malaysia | 62 | 42 | 16 | 42 | 4.38 | 0.59 | 84.2 |
| | AVERAGE | | 21.4 | 15.6 | 59.5 | 7.462 | 1.539 | 74.4 |
| FP | China | 280 (57.1%) | 2 | 1 | 5 | 5.38 | 2.43 | 46.3 |
| FP | Saudi Arabia | 176 (52.1) | 35 | 27 | 56 | 5.97 | 0.46 | 77.2 |
| FP | India | 171 (20.0%) | 11 | 3 | 112 | 3.28 | 0.65 | 61.3 |
| FP | United States | 127 (36.3%) | 1 | 2 | 69 | 16.57 | 3.45 | 79.5 |
| FP | South Korea | 88 (69.3%) | 14 | 9 | 3 | 9.72 | 4.93 | 94.3 |
| FP | Pakistan | 75 (38.9%) | 40 | 30 | 124 | 2.91 | 0.16 | 61.5 |

| | | | | | | | |
|---|---|---|---|---|---|---|---|
| FP | United Kingdom | 66 (40.2%) | 3 | 6 | 34 | 11.34 | 2.91 | 88.8 |
| FP | Italy | 49 (42.2%) | 6 | 8 | 17 | 9 | 1.45 | 91.5 |
| FP | Egypt | 44 (48.4%) | 33 | 36 | 107 | 4.61 | 1.02 | 64.6 |
| FP | Malaysia | 42 (37.5%) | 42 | 16 | 42 | 4.38 | 0.59 | 84.2 |
| | AVER-AGE | | 14 | 10.75 | 52.5 | 8.02125 | 2.055 | 74.9 |

*3.2. Thematic analysis*

According to Zipf bibliometric law [30], 65 most popular author keywords from NFPs and 49 from FPs were taken into VOSViewer analysis. The research landscapes are shown in Figures 2 and 3. SKS performed on those two landscapes resulted in the themes presented in Table 4. SKS analysis of FPs and NFPs author keywords landscapes did not show big thematic differences except two cases. Namely, FPs are more concerned with using AI IoMT applications in e-health and telemedicine, while NFPs are more concerned with using IoMT in pandemic management. Other differences are mainly in the fact that author keywords representing semantically similar topics are differently related, whether funded or unfunded, appear in different topics or differ in popularity.

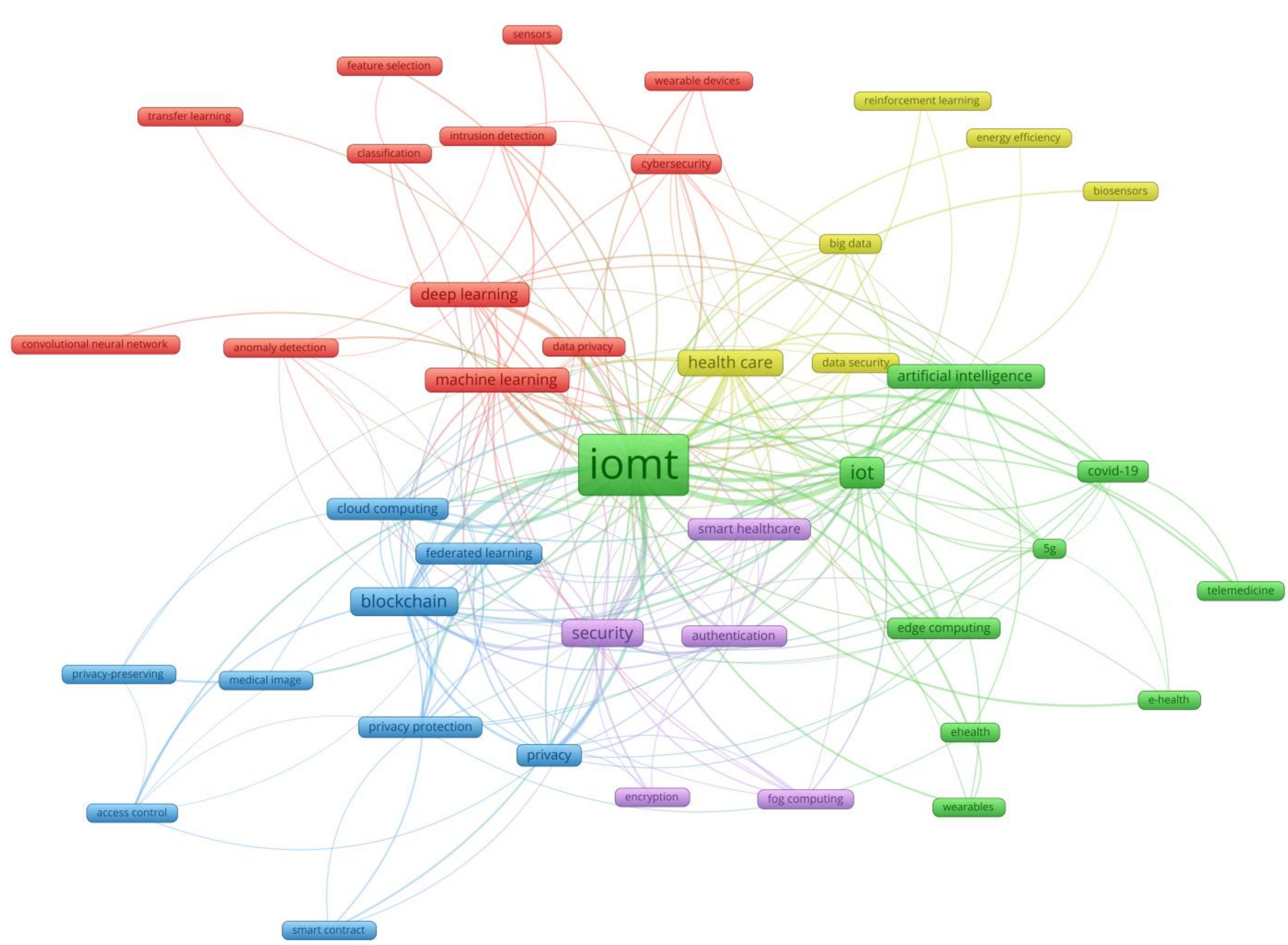

**Figure 2**. The FPs author keywords landscape

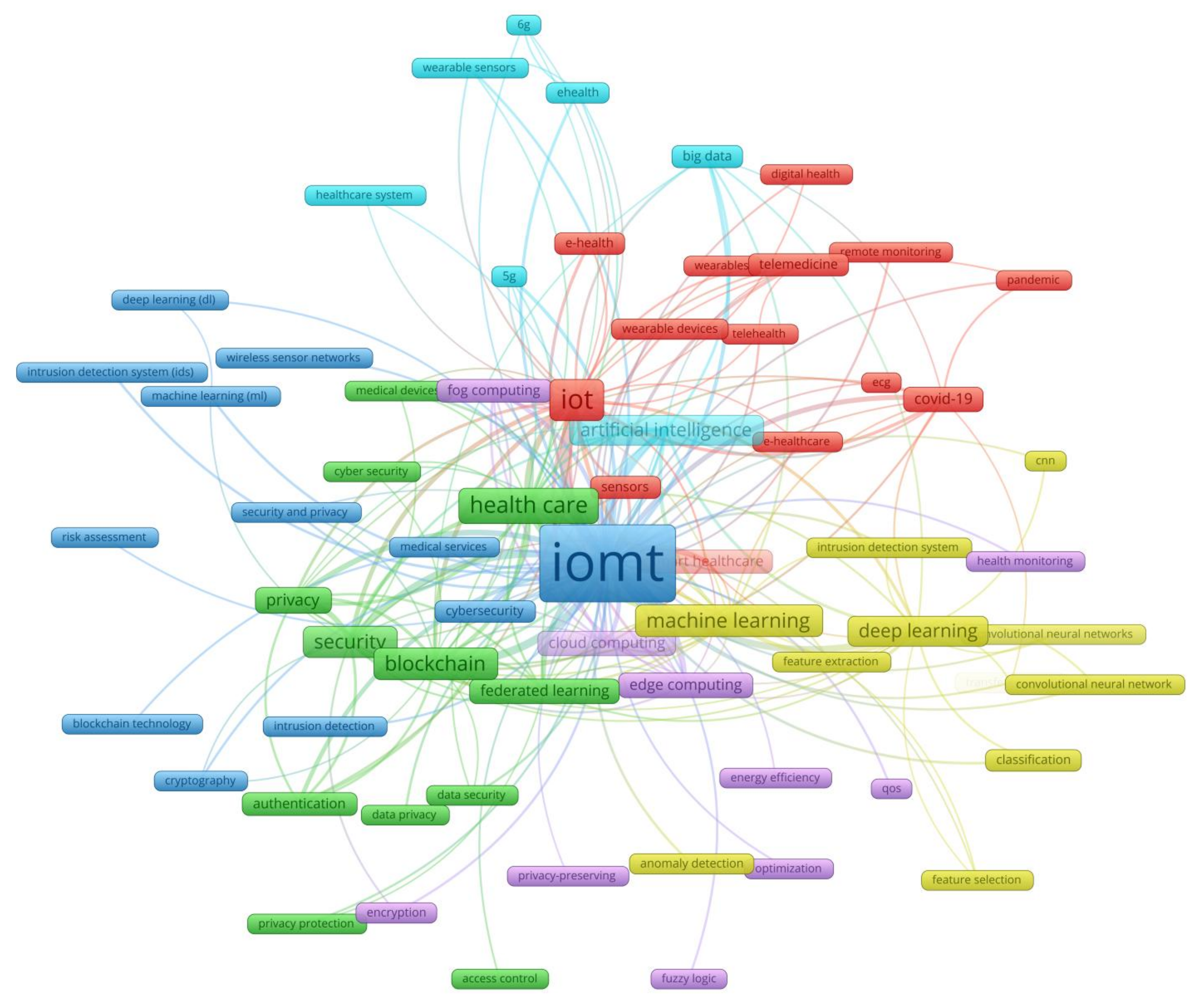

**Figure 3**. The NFPs author keywords landscape

**Table 4.** Themes derived by using SKS

| FPs Themes | Representative topics identified in prominent publications | NFPs themes | Representative topics identified in prominent publications |
|---|---|---|---|
| IoMT and AI use in e-health and telemedicine | ECG monitoring [31], e-health patient monitoring [32], elderly healthcare [33], Accident and emergency detection in One digital health [34] | Role of IoMT in pandemic management | Point of care testing of infectious diseases [35], Cognitive IoMT for pandemic management [36], Pandemic forecasting [37], Covid-19 management by federated learning [38] |
| Privacy in federated learning | Skin diseases [39], Smart healthcare [40], ECG classification [41], Misbehaviour detection [42], Heart disease diagnosing [43] | Privacy and security within federated learning | Privacy preservation with fraud enabled blockchain [44]. Privacy preservation in smart healthcare [45], Intrusion detection [46], Privacy |
| | | | sensitive federated learning [47] |
| Security in smart health care | Blockchain industrial secure encryption in healthcare [48], Hybrid authentication for digital healthcare [49], Threat detection in IoMT networks [50], Secure intelligent biosensors [51] | Machine learning detection of cybersecurity treads on IoMT applications | Cybersecurity of healthcare 5.0 systems using federated learning [52, 53]. Tree classifier based intrusion detection in IoMT [54]. Multilayer perceptron optimisation for cybersecurity [55] |
| Secure big data analysis in healthcare | security threats, vulnerabilities, and counter measures [56], Blockchain, Blockchain assisted big data management [57], Healthcare in Smart Cities [58] | Big data analysis of data from wearable sensors for eHealth | Ambient assisted living [59], edge-stream computing for real time analysis of wearable data [60], big and wearable data in gynaecology [61], Big data based Smart Health Monitoring [62] |
| Advanced machine learning and data security in accessing data from wearables and sensors | Secure wearable ultrasound system [63], Privacy preserving federated learning [64], Robust zero watermarking for federated learning [65], Scalable transferable federated learning in classification of healthcare IoMT data [66] | Advanced machine learning | Remote patient monitoring [67, 68]. Lung tumour diagnosing [69], Digitalization [70] |

### *3.3. IoMT Impact: Bloomberg Global Health Index in relation to the number of funded published papers on IoMT*

Figure 4 represents the BGHI in association with the number of papers published per 1M residents. The association is not strong; however, the trend line (dotted line in Figure 2) shows that the BGHI is increasing with the number of published FPs. The graph is very scattered in the area of less than 0.4 FPs/1M residents, but most of the countries on right of that limit have in general higher BGHI than 80. On the other hand, countries with the lowest BGHIs are all on the left side of the 0.4 FPs/1M residents limit. From the bibliometric point of view the 0.4 FPs/1M residents limit also indicates the point after which the number of FPs do not affect the BGHI in a significant manner. Nevertheless the above patterns might indicate that the investments in IoMT research have positive effect on populational health status.

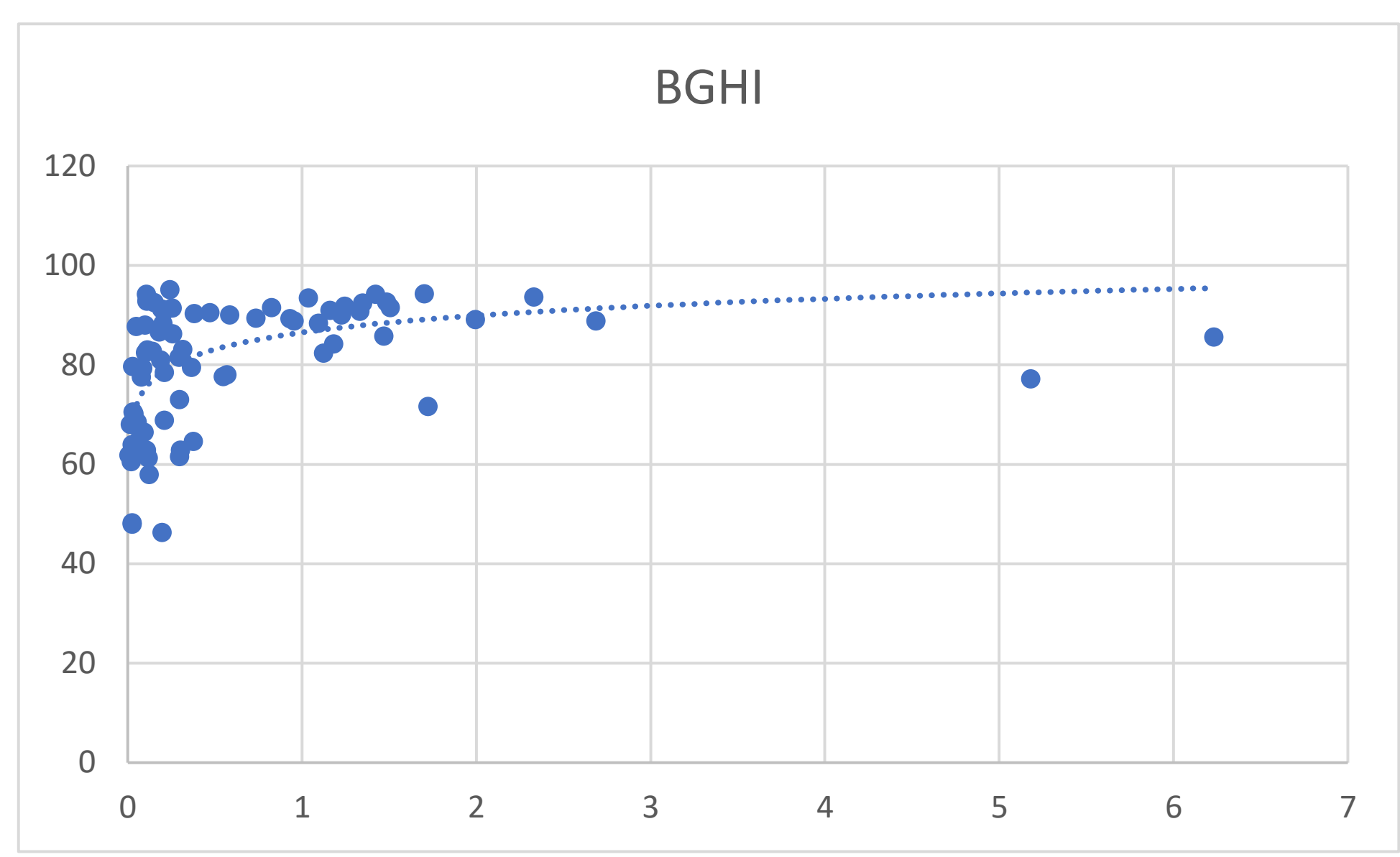

**Figure 4**. The BGHI 2024 in association with the number of published FPs per capita on 1M residents

Most of the most prolific funding agencies (Table 5) funding 13 or more papers (11 agencies) come from China, South Korea, and the EU, which are also among the most productive countries. All of those countries spend a respectable share of GPD on health and research.

**Table 5**. The most prolific funding agencies

| Funding agencie | Number of sponosred publications |
|---|---|
| National Natural Science Foundation of China, China | 47 |
| Conselho Nacional de Desenvolvimento Científico e Tecnológico, Brasil | 29 |
| National Science Foundation, USA | 29 |
| Fundação para a Ciência e a Tecnologia, Portugal | 21 |
| National Key Research and Development Program of China, China | 19 |
| European Comission, EU | 17 |
| Ministry of Science, ICT and Future Planning, South Korea | 17 |
| National Research Foundation of Korea | 17 |
| Ministero dell'Istruzione, dell'Università e della Ricerca, Italy | 16 |
| Horizon 2020 framework, EU | 13 |
| Institute for Information and Communications Technology Promotion, Korea | 13 |

## 4. Discussion

Our study revealed that the proportion of funded projects (FPs) in the Internet of Medical Things (IoMT) research literature was lover than in the fields of IoT in general , and also lower than in the Internet of Things (IoT) research focused on preventive health in particular. This variation in the percentage of FPs can be attributed to several key factors. Firstly, IoT has already a proven potential to revolutionize many disciplines and offers the promise of groundbreaking advancements in a variety of human activities, making it a high-priority area for innovation-driven investment. Similaryl, IoT-based preventive health research, which emphasizes early detection and cost-effective management of chronic diseases, attracts even greater funding. Preventive health solutions using IoT technologies are often viewed as long-term strategies with the potential to reduce healthcare costs and improve population health outcomes, thereby garnering substantial support from stakeholders invested in sustainable healthcare solutions [71]. These approaches are particularly attractive to funders as they help alleviate the burdens on healthcare systems and align with public health initiatives aimed at reducing the prevalence of non-communicable diseases (NCDs). Contrary, IoMT is an emerging field, not yet fully proven [72], making it more likely to attract funding in the future compared to more established fields like general medicine or ICT research. Furthermore, funding agencies may prioritize preventive health due to its proven effectiveness in reducing the incidence of chronic diseases, thereby highlighting the significance of IoT applications in this area [73].

The trend in IoMT research literature production has been positive, reflecting the broader upward trends in literature production across most modern scientific and research fields [74].

In contrast to some other studies [68–70], our analysis did not reveal a regional concentration in research and literature production. The most productive countries in both non-funded projects (NFPs) and funded projects (FPs) include not only well-developed and wealthy nations but also developing countries with less successful economies and healthcare systems. This may be because IoMT-based health solutions can help mitigate workforce shortages in healthcare, address the effects of climate and demographic changes, and improve access to healthcare in remote areas of larger, less developed countries more efficiently than traditional methods [20,71–73]. The lack of regional concentration in IoMT research may also explain why the ranking of the 10 most productive countries in this field differs from their rankings in more general research areas.

Our study also revealed that up to a certain threshold (0.4 FPs per 1 million residents), the BGHI is increasing with the number of funded papers, suggesting that IoMT research funding may contribute to improved healthcare delivery. This finding aligns with other studies that have analyzed the association between research grants and health indices [74–76].

The funding patterns identified in our study can assist researchers in pinpointing suitable research themes for funding, identifying funding institutions, and locating productive countries for potential research collaborations. Additionally, these findings may be valuable to research managers, funding body administrators, government decision-makers, and policymakers.

This article presents a comprehensive analysis of IoMT research, demonstrating both strengths and limitations. A notable strength is the application of the Systematic Knowledge Synthesis (SKS) methodology. This established approach enabled a comprehensive thematic analysis of IoMT research, providing a structured and in-depth understanding of the field. Additionally, this study is the first to meticulously examine funding patterns and the impact of funding within IoMT research, filling a significant gap in the existing literature. A key limitation of the thematic analysis is its reliance on a single database, which may have excluded relevant literature, particularly studies available on various preprint platforms. However, it should be noted that these platforms typically do not include funding information, which was a primary focus of this study. A further limitation arises from the exclusion of "IoMT" as a direct search term to avoid ambiguity with other acronyms such as "Internet of Mobile Things," "Internet of Manufacturing Things," and "Internet of Military Things." This strategic decision, while necessary for precision, may have led to the omission of some pertinent publications.

## 5. Conclusions

The trend of the IoMT research literature production regarding both funded and nonfunded papers is positive, however the percentage of funded papers is decreasing. Government spending in health and research and the health system rank didn’t seem to be associated with research literature production or funding. Nevertheless, funded papers seem to be published in slightly higher ranked journals, and more funded papers are affiliated to scientifically higher ranked countries and country with better country determinants. The research funding expressed with the number funding papers per capita, shows a positive trend with the BGHI.

**Author Contributions:** All authors contributed to the study conception, design and writing of the paper.

**Conflicts of Interest:** The authors have no relevant financial or non-financial interests to disclose.

**Funding:** No funding has been received for research presented in this paper